\documentclass[twocolumn]{aastex63}

\newcommand\rhocas{$\rho$ Cas }
\usepackage{hyperref}

\received{August 26, 2021}
\revised{December 20, 2021}
\accepted{December 21, 2021}

\submitjournal{JAAVSO}

\shorttitle{Exploring the outbursts of \rhocas from visual observations}
\shortauthors{Maravelias G. and Kraus M.}

\graphicspath{{./}{figures/}}

\begin{document}

\title{Bouncing against the Yellow Void - exploring the outbursts of \rhocas from visual observations}

\correspondingauthor{Grigoris Maravelias}
\email{maravelias@noa.gr}

\author[0000-0002-0891-7564]{Grigoris Maravelias}
\affiliation{IAASARS, National Observatory of Athens, GR-15236, Penteli, Greece}
\affiliation{Institute of Astrophysics, FORTH, GR-71110, Heraklion, Greece}
\affiliation{Hellenic Amateur Astronomy Association, GR-10676, Athens, Greece}

\author[0000-0002-4502-6330]{Michaela Kraus}
\affiliation{Astronomical Institute, Czech Academy of Sciences, Fri\v{c}ova 298, 251\,65 Ond\v{r}ejov, Czech Republic\\}

\begin{abstract}

Massive stars are rare but of paramount importance for their immediate environment and their host galaxies. They lose mass from their birth through strong stellar winds up to their spectacular end of their lives as supernovae. The mass loss changes as they evolve and in some phases it becomes episodic or displays outburst activity. One such phase is the Yellow Hypergiants, in which they experience outbursts due to their pulsations and atmosphere instabilities. This is depicted in photometry as a decrease in their apparent magnitude. The object $\rho$ Cassiopeia (Cas) is a bright and well known variable star that has experienced four major outbursts over the last century, with the most recent one detected in 2013. We derived the light curves from both visual and digital observations and we show that with some processing and a small correction ($\sim0.2$ mag) for the visual the two curves match. This highlights the importance of visual observations both because of the accuracy we can obtain and because they fully cover the historic activity (only the last two of the four outbursts are well covered by digital observations) with a homogeneous approach. By fitting the outburst profiles from visual observations we derive the duration of each outburst. We notice a decreasing trend in the duration, as well as shorter intervals between the outbursts. This activity indicates that \rhocas may be preparing to pass to the next evolutionary phase. 

\end{abstract}

\keywords{Massive stars (732) --- Yellow hypergiant stars (1828) --- Stellar mass loss (1613) --- Amateur astronomy (35)}

\section{Introduction} 

Massive stars are very rare: \textit{"for every $20\,M_{\odot}$ star in the Milky Way there are roughly a hundred thousand solar-type stars; for every $100\,M_{\odot}$ star there should be over a million solar-type stars"} \citep{Massey2003}. However, these stars have a significant impact on their immediate environment as well as their host galaxies. They lose mass through their intense stellar winds and they end their lives through spectacular supernovae. This continuous mass loss transfers energy and momentum to the interstellar medium, and it enhances it with material that has been produced in their cores as they evolve. Currently, we are not certain about how a massive star evolves exactly from a main sequence star to more evolved phases and in between them, although we have uncovered many of their properties \footnote{There are a few groups in the world dealing with the details of stellar evolution and their results do not always agree \citep{Martins2013}.}. The main factors that influence stellar evolution and the final stage of (single) massive stars are metallicity, rotation, and mass loss \citep{Ekstrom2012, Georgy2013, Smith2014}. Moreover, the presence of a companion, which seems to be the rule rather than the exception in massive stars ($\sim50-70\%$ in binary systems; \citealt{Sana2012, Sana2013, Dunstall2015}), substantially affects the evolution though strong interaction and mass exchange. The mass loss changes with the  evolutionary phase and in some cases even episodic and/or outburst activity is observed. Examples of such activity are Wolf-Rayet stars, the Luminous Blue Variables, the B[e] Supergiants, the Yellow Hypergiants (YHGs), and the Red Supergiants (RSGs). In most of these cases a complex circumstellar environment is formed, which can be observed as shells, nebulae, or disks (e.g. IRC+ 10420; \citealt{Tiffany2010}).

As the stars evolve beyond the main sequence (ending their hydrogen burning at their cores) they move towards the right part of the Hertzsprung-Russel Diagram (HRD) to the RSGs (burning Helium at their cores). Depending on their mass, rotation, and internal mixing, they may end up as RSGs or, for stars with initial mass range of $\sim20-40\,M_{\odot}$, they can even move back again close to their initial position on the HRD forming a "blue loop". In those cases, there is a region in the HRD ($\sim11000-7000,K$) in which an apparent lack of sources is noticeable. This has been labelled "Yellow Void" \citep{deJager1998}, a temperature regime in which instabilities can occur within the highly inflated envelopes of these objects that might lead, under certain conditions, to eruptions and mass ejections leading to circumstellar shells or envelopes.
During such an outburst the released material obscures the hot atmosphere of the star, which looks fainter and cooler in total. This process may repeat many times up to the point that the largest fraction of the atmosphere is lost and the star is found in a hotter and more stable phase. Then it has passed through the Yellow Void and it appears on the other side as a blue supergiant \citep{Aret2017a, Davies2007}.      

One such star is $\rho$ Cassiopeia (\rhocas; RA: 23:54:23.03 +57:29:57.7), which is a very bright star ($V\sim4.6$ mag) easily spotted in the Cassiopeia constellation. It is a variable star with a modulation determined by multiple long periods \citep{Percy2000} and it has exhibited four outbursts with magnitude drops over 1 mag in 1945-1947 and 2000-2001, and $\sim0.6$ mag in 1985-1986 and 2013-2014 (see \citealt{Kraus2019} and \citealt{Lobel2003} for an overview). In  \cite{Kraus2019} the light curve of CCD photometry obtained from a single observer covering the whole outburst and additional data from the Bright Star Monitor of the American Association of Variable Star Observers (AAVSO) was used in  correlation with spectroscopic observations. In this work, we extend the photometric analysis by using all available digital data, as well as visual observations, which could be found in the AAVSO International Database for the particular outburst. In addition, we explore the whole light curve (from both visual and digital as obtained from a variety of techniques) to study the outburst activity in total. 

Such an investigation requires the longest light curves possible (century long). Ideally these would consist of excellent quality data obtained with the same instrumentation and observing strategy. However, technology changes significantly over this time span and calibration of the different techniques is mandatory to derive robust conclusions. Therefore, we examine the visual observations (that provide the largest coverage) and compare them to the digital ones.

\section{Data collection and analysis} 
\label{s:data_analysis}

We used observations spanning the activity between (March) 1941 to (June) 2021, splitting them into two sets: a. visual observations (magnitude estimates using naked eye or visually
using binoculars or small telescope), b. digital observations. The second set is a collection from various instrumentation (photoelectric photometers, CCD and DSLR cameras), all of which have been reported to the standard Johnson V filter. In general, there can be systematic differences between the various techniques and between observers. However, in the current approach we selected the best observations (marked as non-discrepant by the AAVSO) which simplifies the analysis, as we avoid dealing with these systematics. We also used digital observations retrieved from \cite{Leiker1987} and \cite{Zsoldos1991} for reasons that will become clear later on.

\begin{figure*}
\includegraphics[scale=0.6]{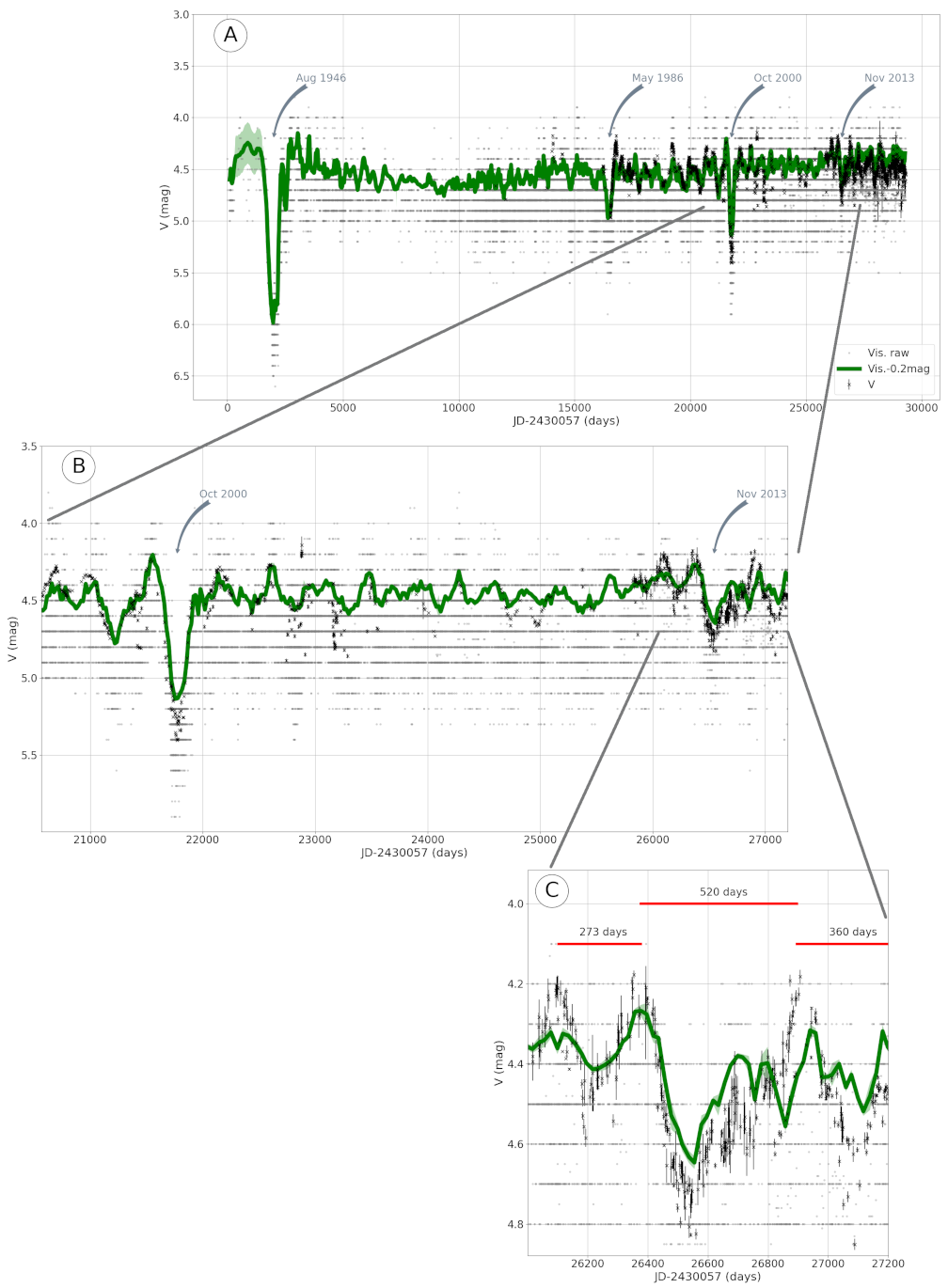}
\caption{\textit{Panel A:} The light curve of \rhocas for the period 1941-2021 (x-axis in Julian Dates - 2430057, corresponding to the initial date). The green line corresponds to the (processed) visual observations (with raw observations shown as gray points) and the shaded area to the 1$\sigma$ error, while digital observations are shown as black dots. After a minor correction of -0.2 mag for the visual light curve the shape of the two curves is almost identical. \textit{Panel B:} Zoom-in of the two major outbursts in November 2000 and 2013. \textit{Panel C:} Zoom-in of the 2013 outburst with various periods (of $\sim300-800$ days; \citealt{Percy2000}) highlighted (see section \ref{s:data_analysis} for more details).} 
\label{f:lightcurve}
\end{figure*}

In the visual approach observers are using standard charts (i.e. selected comparisons stars) that help to reduce systematic differences. However, even in this case the magnitude estimate is based on the perception ability of each individual. Therefore, there is a significant spread of the reported values even for the same epochs (nights). However, with some proper statistical treatment we can obtain a more accurate result. To address that and in order to exclude any short-term variability, we smoothed all data by using a moving average with a window of 30 days (the number was derived by visual inspection of the resulting light curves). Then we grouped the smoothed data into 20-day bins (a typical required frequency of visual observations for long period variables), from which we derive a mean value and its corresponding standard deviation. Finally, we kept only those observations within one standard deviation from the mean value. Starting from 53560 visual observations\footnote{We removed 11 observations with upper limits, i.e. values indicated as "fainter than" with $'<'$.} (from 772 unique observers) we kept 34604 ($\sim65\%$). Using those we re-estimate the mean values and the standard deviations (as errors) at each 20-day bins (1448 points in total). From these data we obtain the (green) light curve, as shown in Fig. \ref{f:lightcurve}, where individual visual observations are shown as gray points (the green shaded area corresponds to the $1\sigma$ error). This includes all visual observations in the last 80 years, from March 3rd, 1941, to June 1st, 2021 (there are about 20 observations omitted before and after these dates). 

In \cite{Kraus2019} measurements from the AAVSO's Bright Star Monitor\footnote{\url{https://www.aavso.org/bright-star-monitor-section}} (514 observations) and from one specific observer (W. Vollmann; AAVSO code VOL; 243 points), who covered the whole 2013 outburst, were used.
In this work, we expanded the coverage of the 2013 outburst with more observations. We also considered all available observations around the outbursts\footnote{Except for the 1946 outburst, for which this technology was simply not available.} and up to June, 2021. In order to improve the coverage of the 1986 outburst the AAVSO data were supplemented by 67 and 83 observations by \cite{Leiker1987} and \cite{Zsoldos1991}, respectively\footnote{There are a few more digital observations since 1960's and prior to this outburst (e.g. \citealt{Brodskaya1966, Landolt1973, ArellanoFerro1985}), but since these data only sample dates outside the outburst, we refrained from adding them into the current work.}. 

Even though they originate from different systems and sensors, the reported magnitudes are given to the standard photometric V filter, providing us with a relatively homogeneous sample. The digital set consists of 2208 measurements in total (2058 from 59 AAVSO unique observers, and 150 from the two papers). A fraction of this data ($\sim10\%$, 222 observations) do not include an error measurement. To estimate it we used a mean error derived from the rest 1986 observations using two approaches: a. a simple average value at 0.014 mag, b. a median value at 0.006 mag. Although the latter is a reasonable error routinely reported in such observations (especially in the more recent CCD and DSLR observations) we opted to use the former value, which is a more conservative approach. The digital observations with their corresponding errors are shown as black x-points in Fig. \ref{f:lightcurve}.

\section{Results and Discussion}

\subsection{Visual vs. digital observations}

In Panel A of Fig. \ref{f:lightcurve} we present the independently extracted light curves from visual (with a -0.2 mag offset) and digital observations. We highlight the four major outbursts that the star has experienced over the last century (henceforth indicated as 1946, 1986, 2000, and 2013). 

A small offset in the visual light curve is necessary in order for the visual light curve to match better the digital one. Then, they become almost identical as it is better shown in the zoom in Panel B of Fig. \ref{f:lightcurve}, where we focus in the last two outbursts of 2000 and 2013 (and where the visual and digital coverage maximizes). This is consistent with the offset of 0.3 mag found by \cite{Percy1985}\footnote{They define as conversion factor between the visual band and the photometric V filter the parameter: $ e = \textrm{(V-Vis)} / \Delta \textrm{(B-V)} \sim 0.25$ mag, where $\Delta \textrm{(B-V)}$ is the difference of the color index for \rhocas (B-V=1.3) with the average color index of comparison/reference stars (B-V=0.1). }. In our case and due to the pre-processing of the visual observations a slightly smaller offset is enough to match the two curves. 
Although for the most part the two curves are identical there are some small noteworthy differences:

\begin{enumerate}
    \item Although supplementing the dataset with more digital observations for the 1986 outburst, we notice that the minimum is still not properly sampled. However, there is a fairly good agreement between digital and visual observations during the recovering phase, with the former pointing to a brighter magnitude than the latter (visual observations have been smoothed, so digital ones are more sensitive to the cyclic behavior).
    
    \item There is a difference in the minimum values of the 2000 outburst ($\textrm{mJD}\sim21775$ - where $\textrm{mJD}$ corresponds to the day from the initial x-axis coordinate set at  $\textrm{JD}-2430057$) and of the 2013 event ($\textrm{mJD}\sim26550$).
    
    \item Digital observations appear fainter than the visual ones at $\textrm{mJD}\sim22800$ and $\sim23200$ with offsets of $\sim0.3$ mag, and at $\textrm{mJD}\sim24736$ and $\sim24976$ with offset of $\sim0.1$ mag.
    
    \item Visual observations appear fainter than the digital ones at $\textrm{mJD}\sim26100$ with an offset of $\sim0.1$ mag.
    
    \item There is a mismatch between $\textrm{mJD}\sim26640-27200$. It is possible that these differences, and especially those at the minima, may be attributed to color changes of \rhocas that are perceived differently between the visual observers (which see all the visual spectrum) and the digital sensors equipped with the photometric filters. Further investigation of the reasons behind this requires a complete set of multi-color observations at these epochs which is not easily available and it is beyond the scope of this paper (an opacity imposed effect based on a gradient in the temperature is discussed in \citealt{vanGenderen2019}).
    
\end{enumerate}

In general, we can conclude that even though the visual approach may seem (and it is) simplistic, when enough data are available they can provide accurate results, almost identical to the ones obtained by digital means, which are more precise but at the expense of more complicated procedures. 

\subsection{Cyclic and outburst activity}

Even from a visual inspection of the light curves in Fig. \ref{f:lightcurve} we can easily spot the cyclic activity of \rhocas outside the outbursts. These periods are of the order of a few hundred days ($\sim300d$, $\sim500d$, and $\sim800d$; \citealt{Percy2000}). In Panel C of Fig. \ref{f:lightcurve} we show a zoom-in of the light curve centered around the 2013 outburst, where we highlight the aforementioned periods. The magnitude drops as measured from the visual and the digital curve are $\sim0.4$ mag and $\sim0.6$ mag, respectively, with an averaged value of 0.5 mag. The profile of the outburst is characterized by a sudden drop in the magnitude (due to the intense release of the material) and a gradual return to the normal state (as the material expands further). 

As the 2013 was the fourth outburst recorded in the history of \rhocas and came a few years after the 2000 one, we were motivated to explore the outburst activity with time. For this we opted to use only the visual observations for the following reasons: a. good match between the visual and digital observations, b. there are no digital observations for the 1946 and 1986 outbursts, while different equipment for digital observations has been used over the years, c. they provide a homogeneous approach as the method has not changed over the years. 

\begin{figure}
\centering
\includegraphics[width=0.5\textwidth]{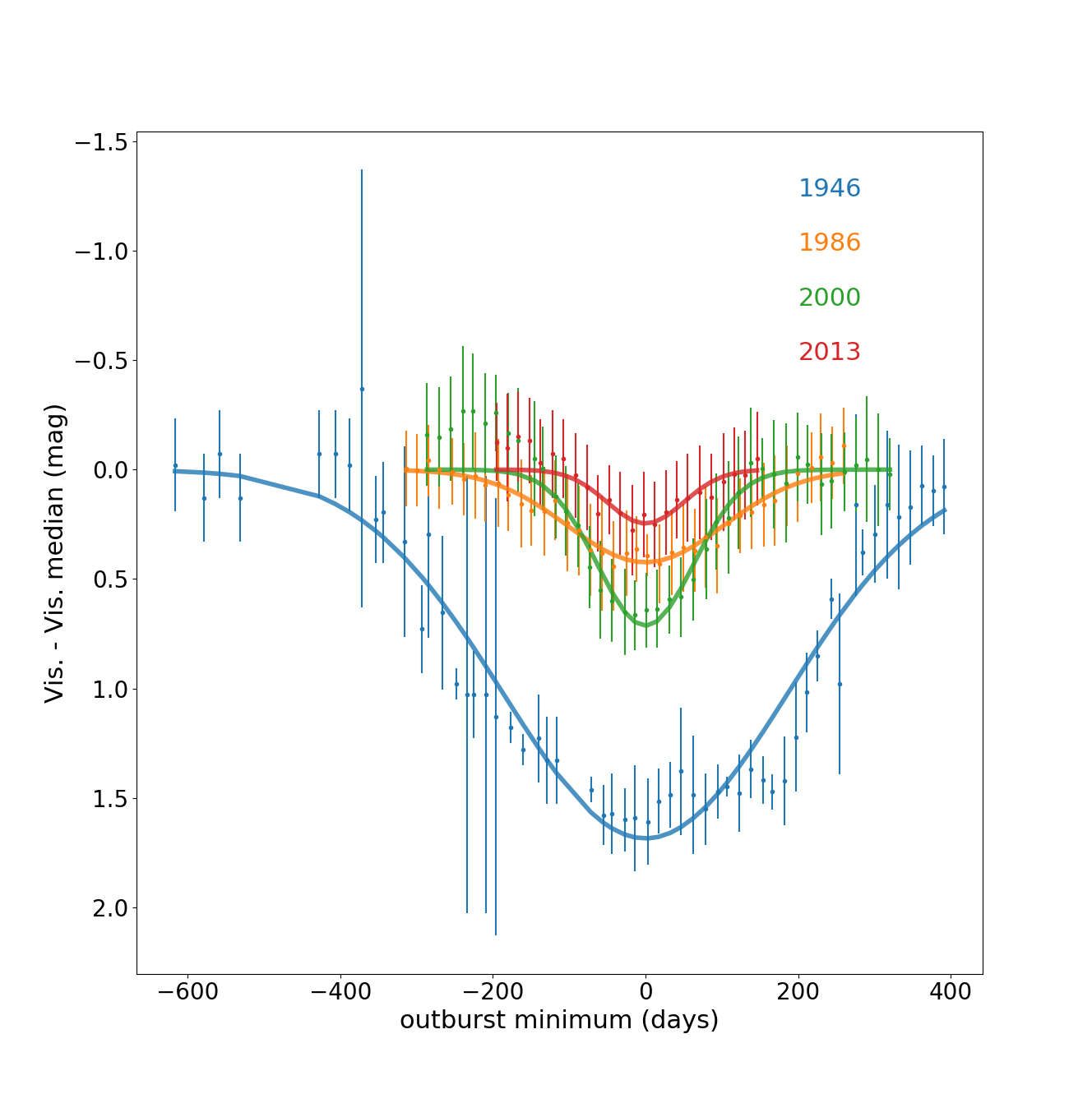}
\caption{Gaussian fits (lines) of the four outbursts of \rhocas. The points and the errors correspond to the moving-average processing of the visual observations for each outburst. All outbursts have been centered to their date of minimum (as identified from the fitting process) so that the x-axis refers to days from outburst minimum. The magnitudes have been normalized with respect to median Vis. magnitude derived from about 1000 days before the star and after the end of each outburst.} 
\label{f:fits}
\end{figure}

For each outburst we first determined a baseline magnitude derived by taking the median from (smoothed) observations about 1000 days before the start and after the end of each outburst (start/end dates were determined visually). Then, for each outburst profile we plot the difference of the smoothed values with the baseline magnitude (see Fig. \ref{f:fits}), in order to bring all profiles at the same scale and to  erase a possible long-term variability of several thousand days that seems to be present (see Panel A of Fig. \ref{f:lightcurve}). However, with this approach we do not remove the cyclic behavior. Obviously, during phases of quiescence the star undergoes cyclic expansion and contraction of its envelope, which sometimes ends up in an outburst. Due to the high luminosity over mass ratio of \rhocas it is plausible that these outbursts could be triggered by pulsational, so-called strange mode instabilities excited in the extended tenuous envelope\footnote{A current theoretical investigation for the stellar parameters of \rhocas confirms the occurrence of very strong strange mode instabilities in this object (Glatzel et al. in preparation).}. After the outburst, the star needs to settle to a new equilibrium state before starting with the next variability cycle, which can have a different length than the one prior to the outburst. But while the outbursts happen during a phase of expansion hence dimming, the outburst itself cannot be considered as being part of an underlying pulsation cycle. Therefore, it is not reasonable to subtract a strict pulsation variability from the light curve underneath the outburst, and we opted for the simplest approach of fitting a Gaussian function to the observed profiles.

For the fitting process we changed the window of the smoothing to 15 days, as this provides more points for fitting without affecting much the end result. 
The free parameters were the amplitude, the standard deviation ($\sigma$), and the mean that corresponds to the minimum date of the outburst. In Fig. \ref{f:fits} we show the final fits obtained for each outburst while in Table \ref{t:stats} we present the derived parameters, along with their corresponding goodness-of-fit as defined from the $\chi_{red}^2$. We approximated the total duration for each outburst by taking the Full Width at 10\% of the Maximum (FWTM), calculated as $\textrm{FWTM}=4.29193 \times \sigma$. 

The values of $\chi_{red}^2$ are not optimal for two reasons: a. a Gaussian profile is not the most appropriate model as the outbursts are not symmetrical, b. the errors related to the observations are probably overestimated (since they correspond to the spread of the visual estimates). Although their mean values can track the true activity of the star the large spread leads to significant errors that propagate through the fitting to the final estimates of each outburst duration. 

Nevertheless, the actual derived numbers for the duration and the amplitudes are consistent with the values from previous works. For the 2000 and 2013 outbursts, durations of $\sim477d$ and $\sim300d$, and amplitudes of $\sim1$ mag and $\sim0.55$ mag are quoted by \cite{Lobel2003} and \cite{Kraus2019}. It is also worth to note that the depth of the 2013 outburst is the shallowest of all, which results simultaneously in the highest uncertainties. Concurrently, the 1946 outburst is the longest and deepest outburst observed so far. 

\begin{table*}[]
    \centering
    \caption{Outburst properties and statistics.}
    \begin{tabular}{ lccccccc } 

     \hline 
     \hline
     Outburst label & Amplitude & $\sigma$ &  JD of Minimum  &  Date of Minimum  & d.o.f.* & $\chi_{red}^2$ & Duration \\
                    &  (mag)    & (days)   & (days)& (DD/MM/YYYY)  &         &                & (days) \\  
     \hline
     1946 & 1.69 $\pm$ 0.27 & 187 $\pm$ 36 & 2432048 $\pm$ 35 & 15/8/1946 & 53 & 1.37 &  800 $\pm$ 154 \\
     1986 & 0.42 $\pm$ 0.36 & 102 $\pm$ 99 & 2446539 $\pm$ 98 & 18/4/1986 & 36 & 0.06 &  437 $\pm$ 426 \\
     2000 & 0.71 $\pm$ 0.45 & 63 $\pm$ 45 & 2451840 $\pm$ 46 & 22/10/2000 & 38 & 0.26 &  270 $\pm$ 196 \\
     2013 & 0.29 $\pm$ 0.50 & 50 $\pm$ 118 & 2456603 $\pm$ 118 & 7/11/2013  & 21 & 0.13 &  214 $\pm$ 509 \\
     \hline
     \hline
     \label{t:stats}
     \end{tabular}
    \textit{ *d.o.f. is the number of smoothed magnitude values per outburst minus the three free parameters}
    
\end{table*}

\begin{figure}
\centering
\includegraphics[width=0.5\textwidth]{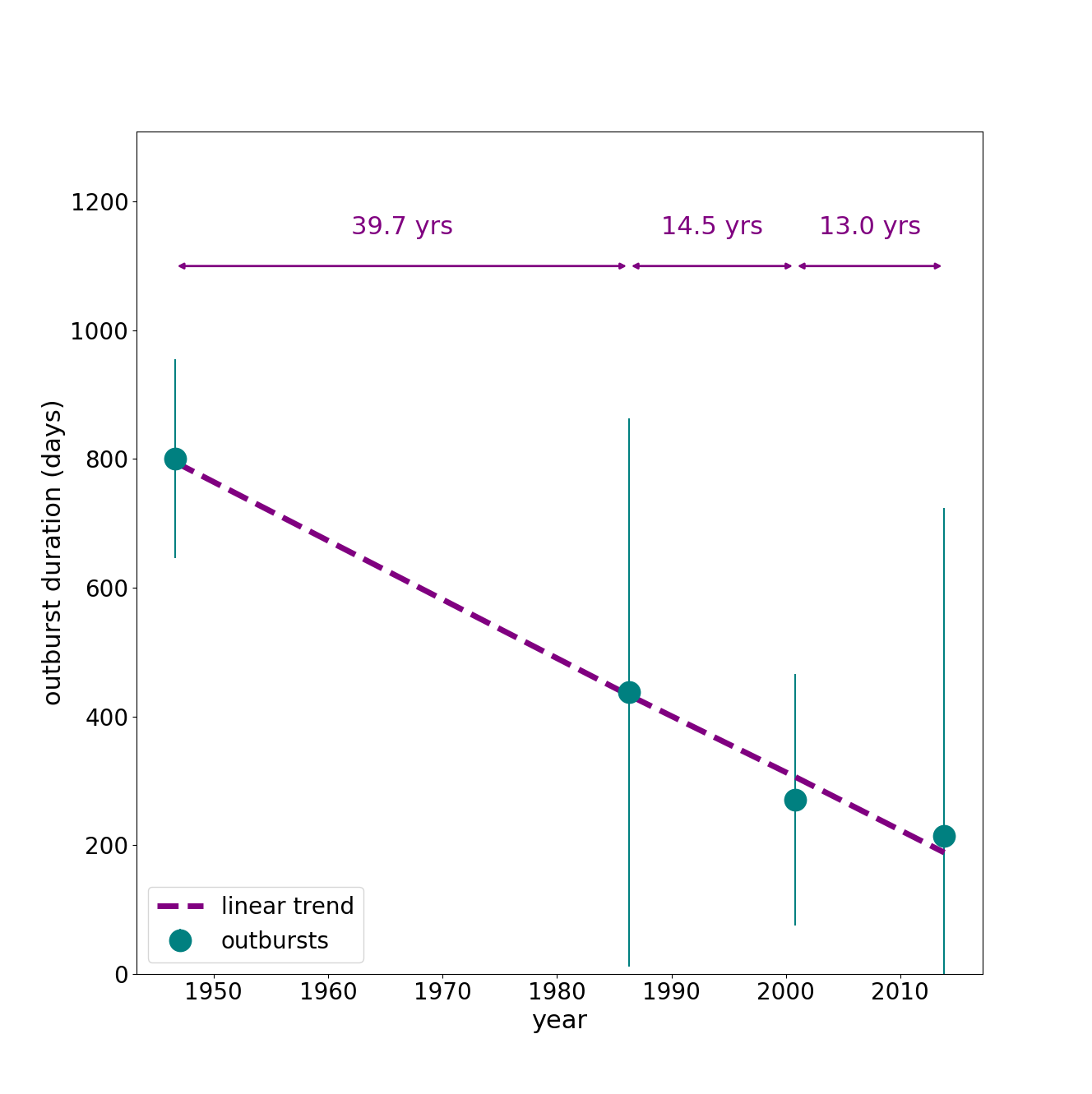}
\caption{The duration of each outburst (dots) with time (using the minimum dates as identified from the fitting process). There is a trend of shorter outbursts with time (linear model indicated with the violet dashed line). They also seem to occur more frequently, as it is indicated by the time difference between the outbursts (violet arrows).} 
\label{f:outbursts_wtime}
\end{figure}

In Fig. \ref{f:outbursts_wtime} we plot the duration of each outburst (dots) with time. What is evident from this plot is that the outbursts seem to become shorter with time. This decreasing trend is also shown with a simple linear fit model (violet dashed line). The slope from this model suggests a shortening trend of approximately -10 days/year. Simultaneously, there is a, relatively, increase in the frequency of these outbursts that occurred at about 40, 15 and 13 years. It seems as if \rhocas is actively "hitting" against the Yellow Void, and possibly preparing to pass through \citep{Lobel2003, Aret2017a}. Although the trend is definitely true some caution should be used with respect to potential extrapolations, based on the model limitations described previously.

As of the moment of writing this article, there is no indication of another outburst yet. We are "only" 8 years since the last outburst, therefore it is more than interesting to keep monitoring the activity of \rhocas in order to catch another one in the (possible) near future\footnote{Such a campaign has been initiated by Ernst Pollmann, see the AAVSO Alert notice 746, at  \url{https://www.aavso.org/aavso-alert-notice-746}.}.

\section{Summary}

\rhocas is one of the brightest and most easily spotted Yellow Hypergiants, with a large set of observations dating almost a century back. Having experienced four major outbursts with the latest one in 2013, only 13 years after the 2000 one, we were motivated to investigate the outburst activity. Only visual observations cover completely the first two outbursts (1946 and 1986). After some processing of the raw visual observations we show that they are a good match (with a small offset of $\sim0.2$ mag) to the digital ones (as shown for the 2000 and 2013 outbursts). Given this result we fit the visual curves for all outbursts to derive their durations and amplitudes. The result is a decreasing trend in duration, i.e. the outbursts become shorter and more frequent. This behavior strengthens the argument that \rhocas is bouncing against the Yellow Void and it is probably preparing to pass through it and transit to a new phase (such as a B[e] supergiant).     

\acknowledgments
GM acknowledges funding support from the European Research Council (ERC) under the European Union’s Horizon 2020 research and innovation programme (Grant agreement No. 772086), and MK from the Czech Science Foundation (GA\,\v{C}R 20-00150S). The Astronomical Institute Ond\v{r}ejov is supported by the project RVO:67985815. We thank Alceste Bonanos for providing feedback on the manuscript. We acknowledge with thanks the variable star observations from the AAVSO International Database contributed by observers worldwide and used in this research. This research has made use of NASA's Astrophysics Data System.

\facility{AAVSO}
\software{NumPy \citep{numpy2020}, 
          Matplotlib \citep{matplotlib}, 
          Astropy \citep{astropy}.
          }

\bibliography{rhoCas}{}
\bibliographystyle{aasjournal}

\end{document}